\documentclass[aps,twocolumn,preprintnumbers,floatfix,nofootinbib]{revtex4}

\usepackage{graphicx}
\usepackage{dcolumn}
\usepackage{bm}
\usepackage{epsfig}
\usepackage{amsmath}
\usepackage{amssymb}

\def\fun#1#2{\lower3.6pt\vbox{\baselineskip0pt\lineskip.9pt
  \ialign{$\mathsurround=0pt#1\hfil##\hfil$\crcr#2\crcr\sim\crcr}}}

\input epsf

\newcommand{\be}{\begin{equation}}
\newcommand{\ee}{\end{equation}}
\newcommand{\bea}{\begin{eqnarray}}
\newcommand{\eea}{\end{eqnarray}}

\begin{document}

\title{From PAMELA to CDMS and back}

\author{Qing-Hong Cao$^{a,b}$, Ian Low$^{b,c}$, Gabe Shaughnessy$^{b,c}$ }
\affiliation{
$^a$ \mbox{Enrico Fermi Institute, University of
Chicago, Chicago, IL 60637}  \\
$^b$ \mbox{High Energy Physics Division, Argonne National Laboratory, Argonne, IL 60439}\\
$^c$ \mbox{Department of Physics and Astronomy, Northwestern University, Evanston, IL 60208} \\
}

\begin{abstract}  
We study implications of the recent results from the CDMS collaboration on astrophysical probes of dark matter.      
By crossing symmetry an elastic scattering cross section with the nucleon implies annihilation of dark matter into hadrons inside the halo, resulting in an anti-proton flux that could be constrained by data from the PAMELA collaboration if one includes a large boost factor necessary to explain the PAMELA excess in the positron fraction. As an illustration,  we present a model-independent analysis for a fermionic dark matter and study the upper bound on the boost factor using the PAMELA anti-proton flux.

\end{abstract}

\pacs{draft}

\maketitle

\noindent {\bf Introduction -- } The nature of the dark matter (DM) is one of most fundamental questions in
contemporary science. One appealing candidate is the weakly interacting massive 
particle (WIMP), which appears in 
many theories beyond the standard model (SM) and naturally gives the observed relic abundance \cite{Bertone:2004pz}.
Since the WIMPs generically interact with the SM particles, experiments have searched for their production in 
collider detectors, scattering off the nuclei in underground laboratories, as well as particles produced by 
their annihilation in the galactic halo. It is widely believed that the WIMPs, if exist,
 must reveal themselves in one of the collider, direct detection, or indirect detection experiments, 
although so far an incontrovertible discovery remains elusive.
 
However, in the last several years we have witnessed several ``anomalies'' in various experiments. In indirect detections,
several collaborations observed excessive fluxes of electrons/positrons and photons, which may not be explained  
easily by conventional astrophysical sources. The experiments  include the PAMELA \cite{Adriani:2008zr},
the ATIC \cite{ATIC:2008zzr}, the Fermi LAT \cite{Abdo:2009zk}, the HESS \cite{Collaboration:2008aaa}, and
the WMAP \cite{Finkbeiner:2004us}. Furthermore, the direct detection experiment DAMA \cite{Bernabei:2008yi} has observed annual modulations
consistent with that expected from dark matter scattering off the nuclei, which nonetheless is difficult to reconcile with null results from
other direct detection experiments in conventional WIMP scenarios.

In the above anomalies, the PAMELA data received perhaps the most attention and sparked a plethora of theories on DM attempting to accommodate
the observed excess in  the positron fraction below 100 GeV \cite{Adriani:2008zr}, which suggests an annihilation cross section $\sigma_{an} v$ in the halo that is 2-3 orders of magnitude larger than the typical WIMP annihilation of roughly 1 pb. 
To achieve such a large cross section, a novel mechanism that is sometimes used is the Sommerfeld enhancement \cite{sommer} in the context of DM annihilations \cite{Hisano:2004ds}.
On the other hand, the PAMELA also observed anti-proton fraction consistent with the expected astrophysical background \cite{Adriani:2008zq}, implying a ``leptophilic" DM annihilating mostly into leptons \cite{Cirelli:2008pk}.

More recently, the CDMS collaboration announced an updated limit on
the DM-nucleon scattering cross section in the vicinity of $10^{-7}$ pb \cite{CDMS}, resulting in a number of works studying its implications~\cite{Kadastik:2009ca}.
However, it appears the parameter space with the potential of giving a positive signal in future 
direct detection experiments such as the XENON \cite{Angle:2007uj} has not been explored in detail in light of the observed anomalies in indirect detection experiments, as most studies focus on evading the limits set by CDMS and XENON.

In this letter we present a model-independent analysis on the implication of a positive signal in DM scattering off nuclei, assuming the DM interacts primarily with quarks inside the nuclei, and take into account a large boost and/or Sommerfeld enhancement factor needed to explain the PAMELA positron fraction.  By crossing symmetry the DM then would have a non-vanishing annihilation cross section into quarks, giving rise to anti-proton fluxes which could be observed by the PAMELA. Our approach is similar in spirit to, but more general than, Ref.~\cite{Barger:2008qd}. As an illustration we focus on the case of a fermionic dark matter, although our analysis could be generalized to other spins easily \cite{longpaper}.


\noindent{\bf Kinematics of DM Detections -- } The kinematic regime of the DM direct detections is 
different from that of the indirect detections. We are interested in relating the elastic scattering cross section per nucleon $\sigma_0$ to the $s$-wave component of the annihilation cross section $\sigma_{an} v$. 
We will assume the dark matter $\chi$ interacts with the quarks through one or more ``mediator'' particles $\phi$ at the tree-level, whose mass is  $m_\phi$. The $\phi$ coupling to the DM is $g_{\chi}$, while its coupling to the quark is  $g_{q}$. 

Consider the scattering 
$\chi (p_\chi) + q (p_q) \to \chi (k_\chi) + q (k_q)$.
In the non-relativistic (NR) limit
we expect the energy of
a bound quark inside the nucleon to be $ < 1\ {\rm GeV} \ll m_\chi$.
The Mandelstam variables are
\bea
s_{el}&=&(p_\chi + p_q)^2 \approx m_\chi^2 \ , \nonumber\\
t_{el}&=& (k_\chi - p_\chi)^2 \approx (m_\chi v)^2\ll m_\chi^2 \ , \\
u_{el}&=&(k_q - p_\chi)^2 \approx  m_\chi^2\ . \nonumber
\eea
By crossing symmetry the amplitude for the annihilation process $\chi (p_\chi) + \chi (k_\chi) \to q (p_q) + q (k_q)$ is obtained from
the scattering by $k_\chi \to -k_\chi$ and $p_q \to -p_q$, as well as the following change of the Mandelstam variables:
\bea
s_{el}&\to& t_{an} = (p_q - p_\chi)^2 \approx -m_\chi^2 \ , \nonumber \\
t_{el}& \to& s_{an} =(p_\chi + k_\chi)^2 \approx 4 m_\chi^2 \ , \\
u_{el} &\to& u_{an}=(k_q - p_\chi)^2 \approx  - m_\chi^2\  . \nonumber
\eea
A simple relation between $\sigma_0$ and $\sigma_{an}v$ exists if  either $m_\phi \agt 2m_\chi$ when there is $s$-channel annihilation
($t$-channel scattering) or  $m_\phi \agt  m_\chi$ without the $s$-channel annihilation. One example is when the DM is lighter than
the $W/Z$ bosons.
In either case both cross sections could be computed by using {\em the same} effective operator. For example, a spin-1 $\phi$ induces the operator $(g_{\chi} g_{q}/m_\phi^2) \bar{\chi}\gamma_\mu \chi \, \bar{q}\gamma^\mu q$ for both $\sigma_0$ and $\sigma_{an}v$. 

Away from the above two scenarios, the effective operator can still be used to compute $\sigma_0$, barring a resonance effect which is model-dependent. However, to compute the $\sigma_{an}v$ for final states with quarks, we need to multiply the DM annihilation cross section into two on-shell $\phi$ particles by the branching ratio into quarks. This happens if the DM can annihilate into the $Z$/Higgs bosons.


\noindent{\bf Dynamics of DM Detections --} We first consider a heavy mediator particle. 
There is only one effective operator contributing to both the spin-independent scattering and the $s$-wave annihilation cross sections at the tree-level \cite{Barger:2008qd}, which has the vector coupling $(\bar{\chi}\gamma^\mu \chi)(\bar{q}\gamma_\mu q)$.

The spin-independent cross section of elastic WIMP-nucleon scattering can be written as
\be
\sigma_0 = \frac{4}{\pi} m_{r}^2 \frac{[ Z f_p + (A-Z) f_n]^2}{A^2}\ ,
\ee
where $m_r= m_\chi m_p/(m_\chi+m_p)\approx m_p$ is the reduced mass of the WIMP-proton system and the same as the proton mass $m_p$ for $m_\chi \gg m_p$.
$A$ and $Z$ are the atomic mass and atomic number of the target nuclei, respectively, and $f_{n,p}$ are the effective couplings of the WIMP to the proton and the neutron. For an effective vector coupling 
\be
\label{eq:vectorop}
\frac{G_V^q}{\sqrt{2}} \bar{\chi}\gamma_{\mu} \chi \bar{q}\gamma^{\mu}q \,
\ee
only the time component of the current $\bar{q}\gamma^\mu q$ is important in the NR limit, which becomes the number density.
Therefore we could write
\be
f_{p}= 2\frac{G_V^u}{\sqrt{2}} + \frac{G_V^d}{\sqrt{2}} \ , \quad f_{n}= \frac{G_V^u}{\sqrt{2}} + 2\frac{G_V^d}{\sqrt{2}}  \ .
\ee
Assuming a universal coupling $G_V^q\equiv G_V$ we arrive at
\be 
\sigma_0 = 2.23\times 10^{-5}~{\rm pb} 
\left[\frac{G_V}{10^{-7}\  {\rm GeV^{-2}}}\right]^2\ .
\label{eq:fermion-direct}
\ee

On the other hand, if the same effective operator in Eq.~(\ref{eq:vectorop}) can be used to describe the DM annihilation in the halo, the cross section into the quark becomes
\be
\sigma_{an}v=\frac{1}{4\pi} \sum_{q} G_{V}^{2} C_F m_{\chi}^2
\sqrt{1-\frac{m_q^2}{m_\chi^2}}\left(2+\frac{m_q^2}{m_\chi^2}\right) \ ,
\label{eq:fermion-ann}
\ee
where $C_F=3$ is the color factor for quarks.  
Substituting $G_V$ in
Eq.~(\ref{eq:fermion-ann}), we obtain the relation 
\begin{eqnarray}
\sigma_{an}v
& = &
1.6 \ {\rm pb} \left[\frac{\sigma_{0}}{4\times 10^{-7}~{\rm pb}}\right] \left(\frac{m_\chi}{\rm TeV}\right)^2 \left(\frac{B}{100}\right) \nonumber \\
&&\quad \times \sum_q\sqrt{1-\frac{m_q^2}{m_\chi^2}}\left(2+\frac{m_q^2}{m_\chi^2}\right) \ ,
\label{eq:fermion-corr}    
\end{eqnarray}
where the kinematic factor in the second line involves summing over all six quark flavors (under the assumption of universal couplings).
In Eq.~(\ref{eq:fermion-corr}) we have also included a boost/Sommerfeld enhancement factor $B$ which is needed to explain the excess in the positron fraction in the PAMELA data. 
Eq.~(\ref{eq:fermion-corr}) then yields a prediction for the anti-proton flux for WIMP annihilations from
the potentially positive signal at the CDMS.

Next we consider a light mediator particle. In this situation the effective operator
in Eq.~(\ref{eq:vectorop}) can still be used to compute $\sigma_0$,
although it is no longer a valid description
for the WIMP annihilation in the halo since $\phi$ could be produced on-shell
from the annihilation and the rate of producing quarks is determined by the branching ratio, which
does not enter into the direct detection.

For a fermionic DM $\phi$ could be either spin-0 or spin-1. However, a scalar $\phi$ suffers from the $p$-wave suppression in the annihilation, 
as mentioned previously, so we will consider a vector mediator using the generic notation $Z^\prime$, which could very well 
be the SM $Z$ boson. Furthermore, we concentrate on the DM being a Dirac fermion. The case of a Majorana fermion can be considered
 in a straightforward manner.

Let's first consider the scenario where the $Z^\prime$ is not charged under the SM strong interaction. The fundamental interactions at the renormalizable level are
\be
g_\chi Z_\mu^\prime \bar{\chi}\gamma^\mu \chi  \quad {\rm and} \quad g_q Z_\mu^\prime \bar{q}\gamma^\mu q \ .
\ee
The WIMP-nucleon elastic scattering occurs in the $t$-channel and the effective operator has the coefficient
\be
\label{eq:gvlight}
\frac{G_V}{\sqrt{2}} = \frac{g_\chi g_q}{m_{Z^\prime}^2}.
\ee
For the WIMP annihilation there are two possibilities:  the $s$-channel into quarks through the $Z^\prime$ exchange and the $t$-channel into
two on-shell $Z^\prime$ bosons, which subsequently decay into the quark according to the branching ratio. 
The $s$-channel cross section is obtained by using
$G_V/\sqrt{2}= g_\chi g_q/(s-m_{Z^\prime}^2)$ in Eq.~(\ref{eq:fermion-ann}), resulting in
\bea
\left.\sigma_{an}v\right|_s 
& = & 
1.6 \ {\rm pb} \left[\frac{\sigma_{0}}{4\times 10^{-7}~{\rm pb}}\right] \left(\frac{m_\chi}{\rm TeV}\right)^2 \left(\frac{B}{100}\right) \nonumber \\
&& \times \frac{r_{Z^\prime}^4}{(4-r_{Z^\prime}^2)^2} \sum_q\sqrt{1-r_q^2}\left(2+r_q^2\right) \ ,
\label{eq:fermion-corr1}    
\eea
where $r_{Z^\prime, q} = m_{Z^\prime, q}/m_\chi$. Notice that $r^4/(4-r^2)^2 \ll 1$ for $r \alt 1$.
On the other hand, 
the cross section for 
the $t$-channel process 
$\bar{\chi} \chi \to Z^\prime Z^\prime$  is %
\be
\label{eq:sanzp}
\left.\sigma_{an} v\right|_{Z^\prime} =\frac{g_\chi^4}{4\pi}\frac{1}{m_\chi^2} 
\frac{(1-r_{Z^\prime}^2)^{3/2}}{(2-r_{Z^\prime}^2)^2}  \  .
\ee
Hence, the annihilation cross section to quarks is 
\bea
\left. \sigma_{an} v \right|_t&=& \left.\sigma_{an} v\right|_{Z^\prime}
 \times {\rm Br}(Z^\prime \to q\bar{q}) {\rm Br}(Z^\prime \to q^\prime\bar{q}^{\prime}) .
\label{eq:ferm-ann2}
\eea
From Eq.~(\ref{eq:gvlight}) we can express
$g_\chi$ as
\be
g_\chi^2 = \frac{1}{g_q^2}
\left[\frac{\sigma_{0}}{4.46\times 10^{-3}~{\rm pb}}\right]
\left(\frac{m_{Z^\prime}}{\rm TeV}\right)^4. 
\label{eq:ferm-gv2}
\ee
Substituting Eq.~(\ref{eq:ferm-gv2}) into Eq.~(\ref{eq:ferm-ann2}) we obtain
\bea
\label{eq:tannnum}
 \left.\sigma_{\rm an} v \right|_t &=&
2\times 10^{-10}~{\rm pb} \left(\frac{0.5}{g_q}\right)^4
\left(\frac{m_{Z^\prime}}{\rm 100~{\rm GeV}}\right)^6  \\
&\times & \left[\frac{\sigma_{0}}{4\times 10^{-7}~{\rm pb}}\right]^2 \nonumber
\frac{r_{Z^\prime}^2 (1-r_{Z^\prime}^2)^{3/2}}{(2-r_{Z^\prime}^2)^2} \left(\frac{B_{eff}}{50}\right) \ ,
\eea
where the effective boost factor $B_{eff}=B \times  {\rm Br}\times {\rm Br^\prime}$.
The function $F(r)=r^2 (1-r^2)^{3/2}/(2-r^2)^2 \sim 0.01 - 0.1$ for $ 0.2 < r < 0.9$. 
Comparing Eq.~(\ref{eq:fermion-corr1}) with Eq.~(\ref{eq:tannnum}) we see the $s$-channel annihilation always dominates over the $t$-channel.

A second scenario is when the $Z^\prime$ carries $SU(3)$ color charge. 
This ``leptoquark'' scenario is less common in the literature on WIMP model-building. Therefore we leave the discussion for a future publication \cite{longpaper}.

\begin{figure}[t]
\includegraphics[scale=0.45]{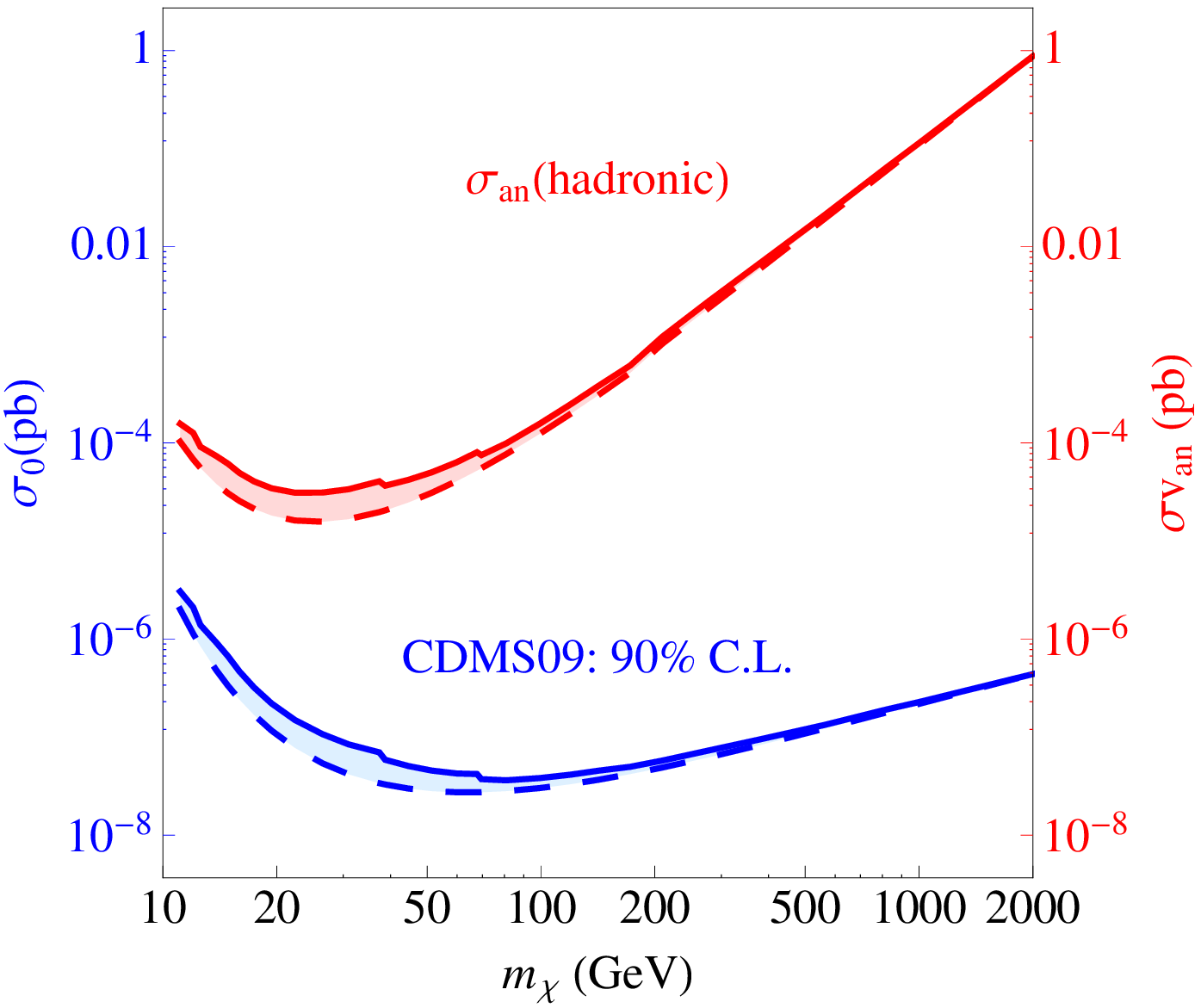}
\caption{CDMS exclusion limit at 90\% C.L. (the left axis) and the inferred DM annihilation into quarks {\em without} the boost factor (the right axis).  For proof of concept, we assume a DM with cross section and mass coverage in the blue region below the present observed limit (blue solid line) and above the expected limit (blue dashed line). The plot assumes a heavy mediator case.
\label{fig:cdmsexcl}}
\end{figure}  

\begin{figure}[t]
\includegraphics[scale=0.44]{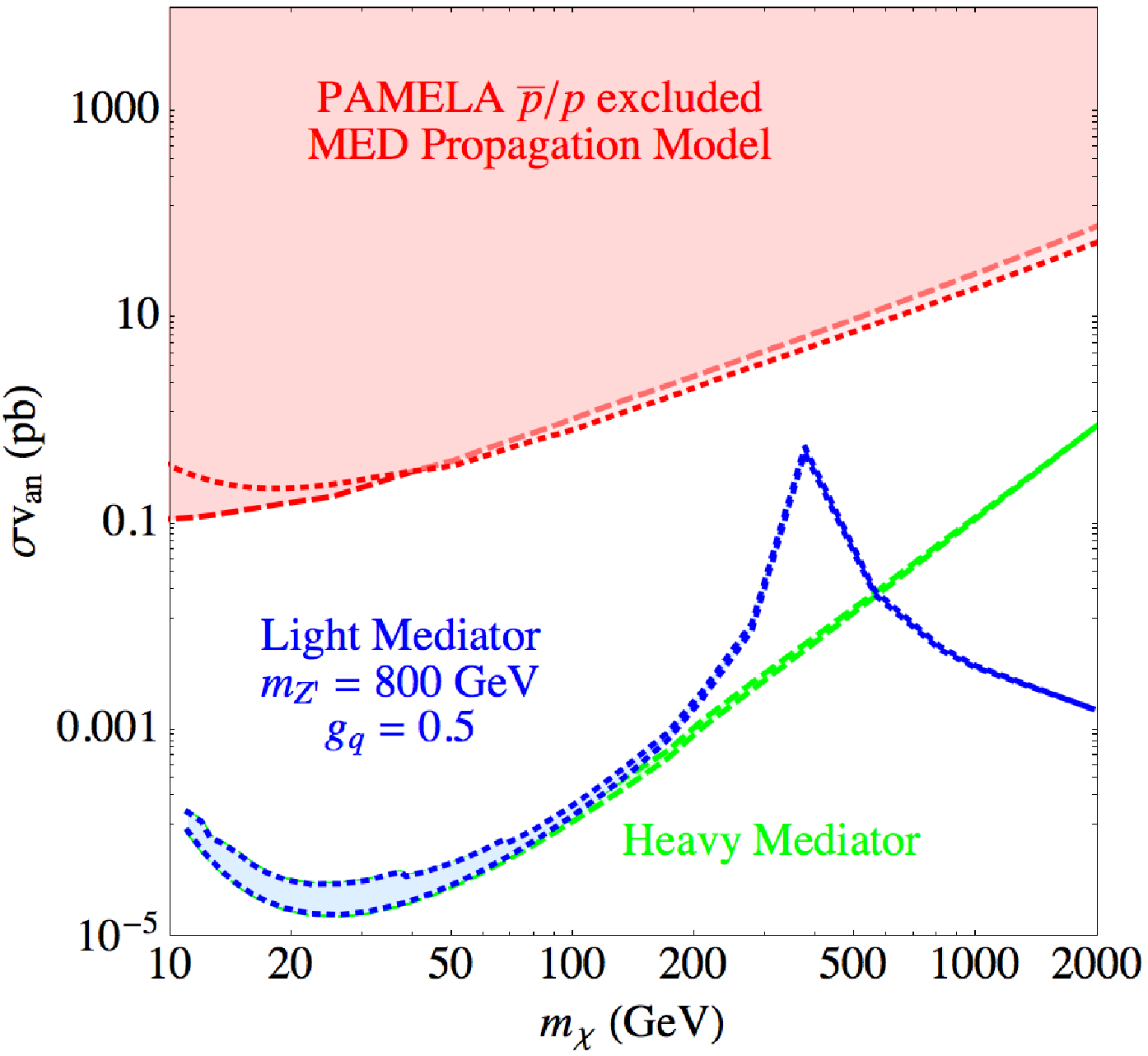}
\caption{The exclusion limits by the PAMELA anti-proton fraction at the 90\% C.~L. The red dashed line is the limit for DM annihilation into two quarks while the red dotted line is for annihilating into four quarks. Again the CDMS inferred annihilation rates are {\em without} the boost factor.
\label{fig:sigv-limits}}
\end{figure}


\noindent{\bf The Result --}
If the recent CDMS observation is a hint that direct detection is ``around the corner'', the analysis in  the previous section suggests a lower bound on the WIMP annihilation into quarks in the halo, giving rise to a substantial anti-proton flux if a large boost factor is included. Since the PAMELA also measured the anti-proton fraction and sees no significant excess below 100 GeV, we could use the anti-proton data to place an upper bound on the boost factor.  For proof of concept for this connection, we assume the DM has a spin-independent elastic scattering cross section that is just below the observed 90\% C.~L. of the latest CDMS results \cite{CDMS}, and consider the resulting the bound on the (effective) boost factor from the anti-proton fraction. Fig.~\ref{fig:cdmsexcl} shows both the bound on the DM elastic scattering cross section from the CDMS collaboration and the inferred $s$-wave component of the DM annihilation into quarks without the boost factor for the case of a heavy mediator.

\begin{figure}[t]
\includegraphics[scale=0.44]{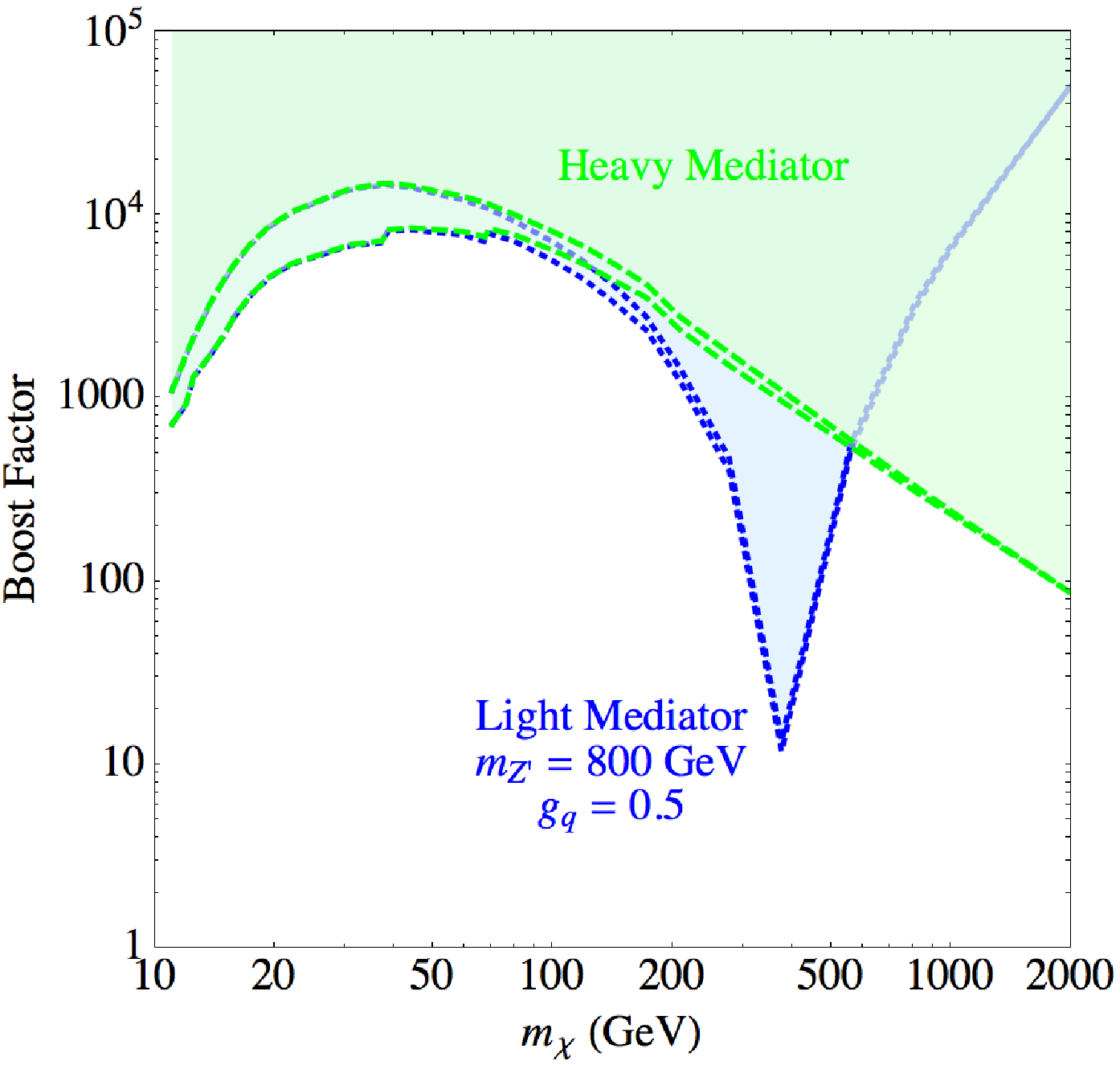} 
\caption{The upper bound on the (effective) boost factor.  The shaded region is disfavored.
\label{fig:dump}}
\end{figure}

To compute the anti-proton flux
we use a halo propagation model that is the med model given in Ref.~\cite{Delahaye:2007fr} and refer the readers there for details. 
In Fig.~\ref{fig:sigv-limits} we show the 90\% C.~L. exclusion limit on the DM annihilation cross section into quarks, as a function of the DM mass, along with the cross sections inferred from the CDMS data. We plot both cases of a heavy and a light mediator. The ratio of the excluded cross section over the inferred cross section gives an upper bound on the boost factor,
which is shown in Fig.~\ref{fig:dump}. One important comment is our knowledge of the background is not perfect, and there are large
uncertainties. In this work we consider the injected primary spectra used in Ref.~\cite{Cirelli:2008id}.

\begin{figure}[t]
\includegraphics[scale=0.44]{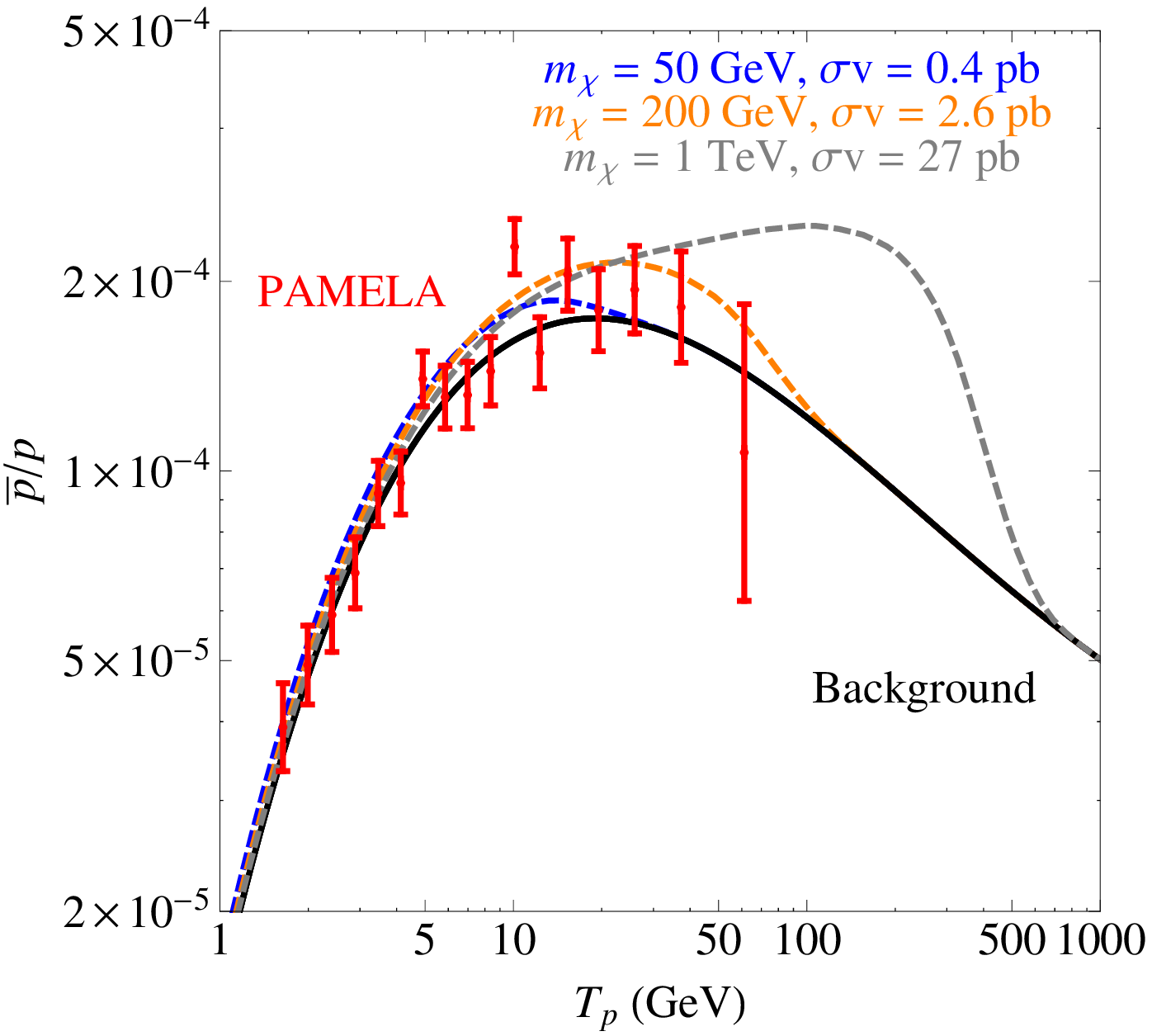} 
\caption{The spectra for the anti-proton fraction, assuming DM annihilation into two quark final states.
\label{fig:ppbar}}
\end{figure}

In Fig.~\ref{fig:sigv-limits} we also show the CDMS inferred annihilation cross section into quarks for both a heavy mediator and a 800 GeV light mediator with $g_q=0.5$, which is slightly smaller than the SM $g_2\sim 0.65$. 
Our benchmark point, a 800~GeV $Z^\prime$ boson, is consistent with the current Tevatron limit of 
the $Z^\prime$ boson in the di-jet mode~\cite{Aaltonen:2008dn}.
Furthermore, such a $Z^\prime$ boson would still belong to the light mediator scenario  as
the dark matter has to be heavier than 2-3~TeV when taking into account of the Fermi LAT results~\cite{Meade:2009iu}.
For the light mediator case we combine both the $s$-channel (into two quarks) and the $t$-channel (into four quarks) results. The PAMELA 90\% C.~L. exclusion limits are insignificantly different for annihilating into two quarks versus four quarks. Taking the ratio of the PAMELA exclusion limit over the CDMS inferred cross section gives an upper bound on the (effective) boost factor allowed so as to be consistent with the anti-proton fraction measured by the PAMELA. The bound on the boost factor is shown in Fig.~\ref{fig:dump}, which in general is consistent with the one needed to explain the 
 PAMELA excess in the positron fraction \cite{Cirelli:2008pk}. One particular region of interest is for $m_\chi \agt 1$ TeV, which is favored by the recent Fermi-LAT measurements on the $e^- + e^+$ spectrum \cite{Abdo:2009zk}, as was argued in Ref.~\cite{Meade:2009iu}.
From Fig.~\ref{fig:dump} we see in this mass range $B\alt {\cal O}(100)$, which is in the lower end of the boost factor required for positron fraction, assuming a heavy mediator case.

In Fig.~\ref{fig:ppbar} we show three different spectra for the anti-proton fraction comparing with the PAMELA data. We choose three sample $(m_\chi, \sigma_{an}v)$ from Fig.~\ref{fig:sigv-limits} to demonstrate the fit to data.

To conclude, we emphasize the connection between the cross section for the elastic scattering off the nuclei measured in  DM direct detection experiments and that of the DM annihilation into hadrons measured by the indirect detection experiments is completely general. If 
the near-future DM direct detection experiments observed a significant number of signal events, 
our analysis could be applied to give a suitable boost-factor constraint.

\noindent {\bf Acknowledgements --}
This work was supported in part by the U.S. Department of Energy under grant numbers
DE-AC02-06CH11357, and DE-FG02-91ER40684.  Q.~H.~C. is supported in part by the Argonne National Laboratory and University of Chicago Joint Theory Institute (JTI) Grant 03921-07-137,
and by the U.S.~Department of Energy under Grants No.~DE-AC02-06CH11357
and No.~DE-FG02-90ER40560. We thank H. Zhang for collaboration in the early stages of this work.
We also wish to thank the organizers and the participants at the Chicagoland Theory Hobnob for a lively
atmosphere where this work was initiated.

\end{document}